\documentclass[fleqn,12pt,twoside]{article}
\usepackage{espcrc1}

\usepackage{graphicx}
\usepackage[figuresright]{rotating}

\newcommand{\AmS}{{\protect\the\textfont2
  A\kern-.1667em\lower.5ex\hbox{M}\kern-.125emS}}

\title{``Point-form'' estimate of the pion form factor revisited}

\author{B. Desplanques\address{Laboratoire de Physique Subatomique 
et de Cosmologie \\(UMR CNRS/IN2P3-UJF-INPG),  
F-38026 Grenoble Cedex, France}\thanks{desplanq@lpsc.in2p3.fr}, 
A. Amghar\address{Facult\'e des Sciences, Universit\'e de Boumerdes, 
35000 Boumerdes, Algeria}, 
and 
L. Theu{\ss}l\address{Departamento de Fisica Teorica, 
Universidad de Valencia,
 \\E-46100 Burjassot (Valencia), Spain}\thanks{Lukas.Theussl@uv.es}   }

\begin{document}

\maketitle

\begin{abstract}
The pion form factor calculation in the ``point-form'' of relativistic 
quantum mechanics is re-considered. Particular attention is given to the 
relation between the momentum of the system and the momentum transfer 
as well as to the quark current.
\end{abstract}

\section{INTRODUCTION}
The charge pion form factor has been considered in many frameworks, 
ranging from field theory to relativistic quantum mechanics ones. 
The latter approaches include well known instant and front forms 
proposed by Dirac \cite{DIRA}. It is relatively recently that the 
point form has been used \cite{ALLE}. At first sight, the implementation 
of this approach agrees well with experiment, suggesting that 
it is on a safe ground.   

It has however been noticed that the same implementation of 
the point-form approach 
(denoted with quotation marks in the following, see comment 
in Ref. \cite{AMGH2}) 
was failing to reproduce the results 
expected from a theoretical model \cite{DESP}. This one, based on scalar 
constituents, is perhaps the simplest one that can be imagined.  
Moreover, the effect of some ingredients, like  form factors 
of the constituents that can affect the comparison of theory 
to experiment, factors out and can be ignored here. 
The theoretical model therefore provides a stringent test 
of the method. It was thus found that the form factors 
obtained in the ``point form'' tend to decrease too rapidly, 
especially in cases where a relativistic approach is required 
(high momentum transfer and/or large binding). The instant 
and front forms do relatively well for standard kinematics \cite{AMGH1}. 

Motivated by the above results, we re-examine the calculation 
of the pion form factor in the ``point form'' of relativistic 
quantum mechanics \cite{AMGH2}. We pay a particular attention to a rescaling 
of the relation  of the Breit-frame pion momentum to the 
momentum transfer, which was made, according to the authors,  
to ensure the correct nonrelativistic limit \cite{ALLE}.
Actually, as the approach is a relativistic one, this modification 
is not needed. Moreover, we include the correct expression for the 
quark matrix element of the current, previously taken as equal to 1.

\section{EXPRESSIONS AND RESULTS}
\begin{figure}[htb]
\begin{center}
\mbox{\includegraphics[width=0.46\textwidth]{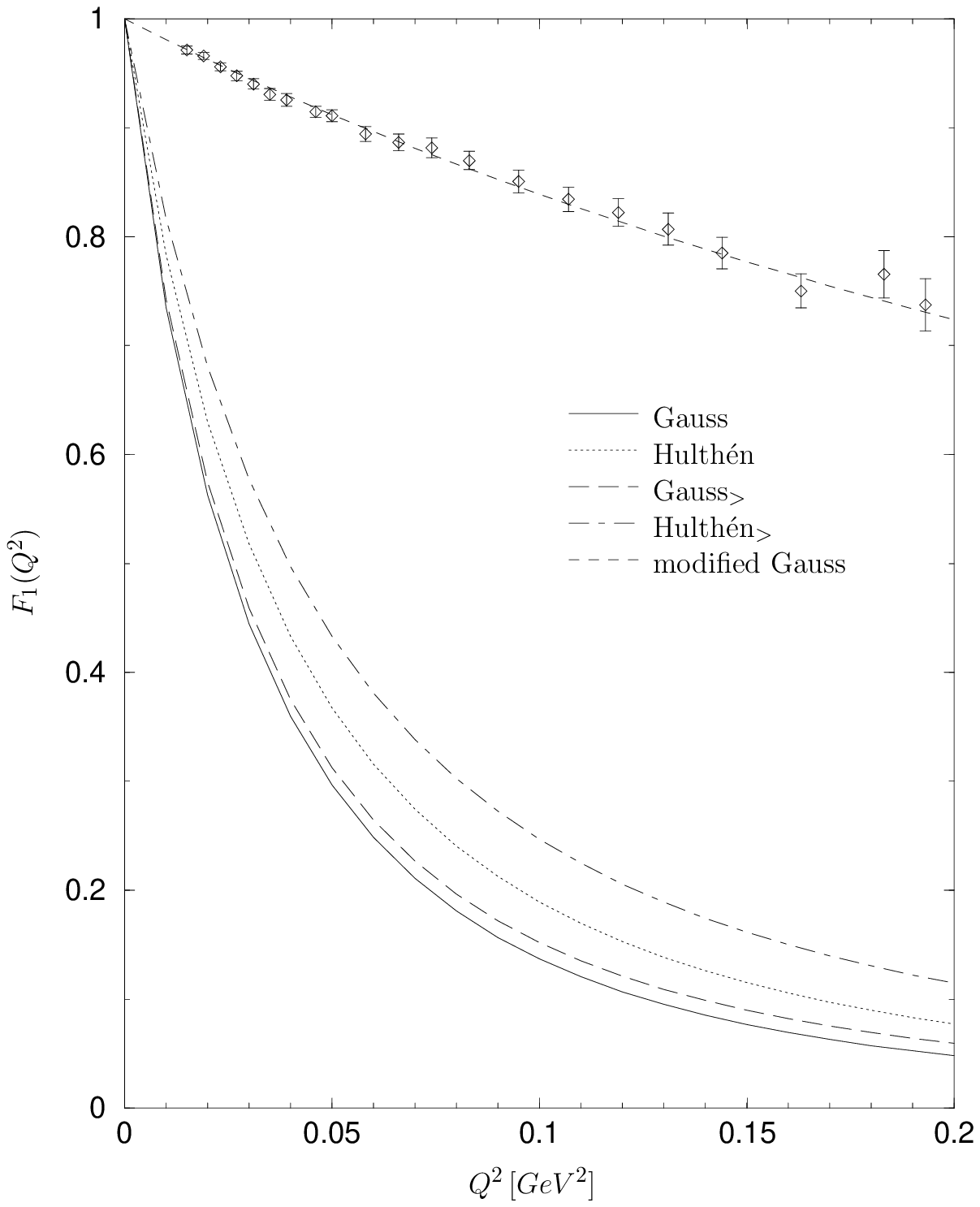}
\includegraphics[width=0.48\textwidth]{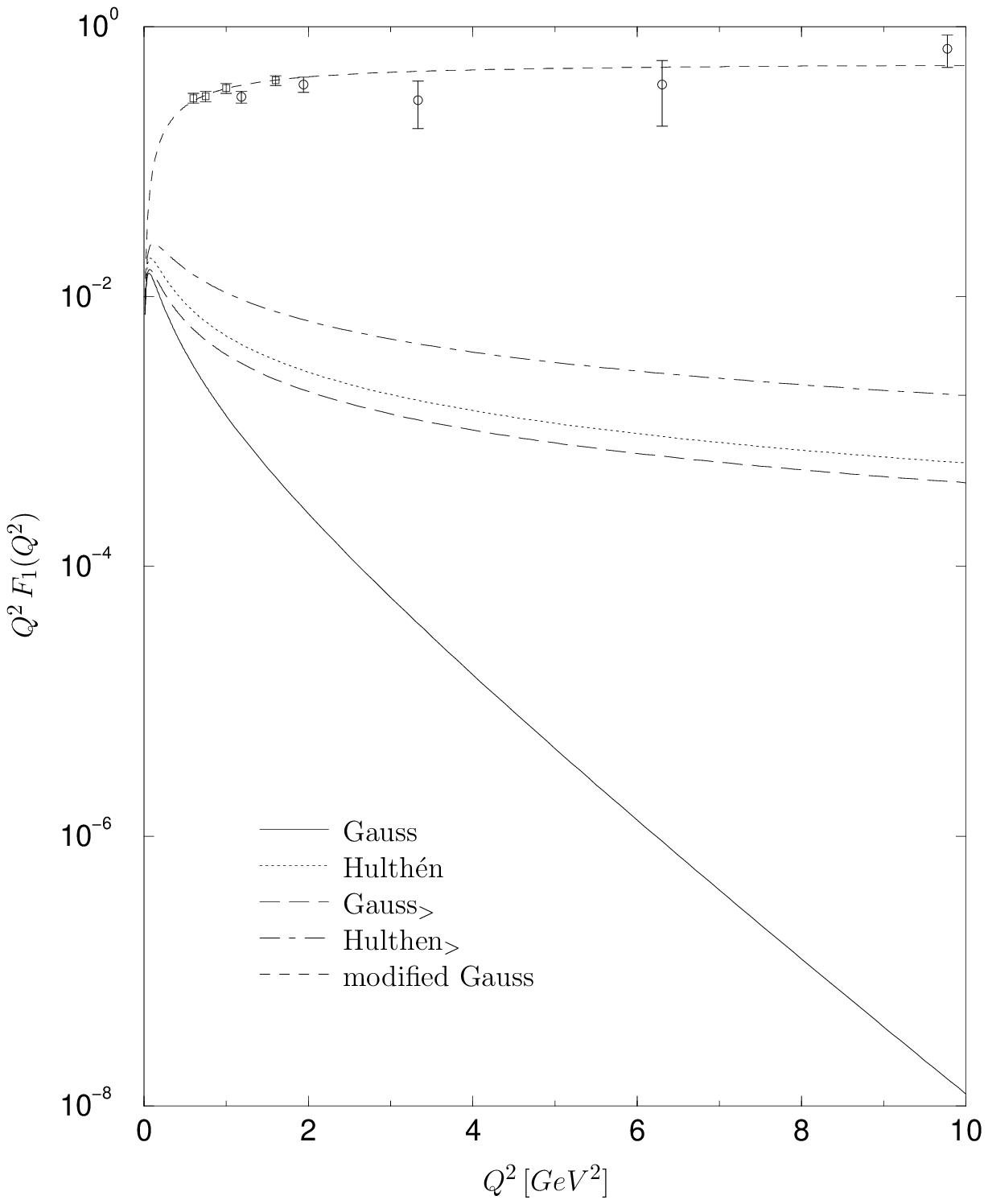}}
\vspace{-6mm}
\end{center}
\caption{Pion form factor calculated with different wave functions in the 
``point form'' of relativistic quantum mechanics. The left-hand graph 
shows the low-$Q^2$ behavior in linear scale, the graph on the 
right-hand side shows the form factors, multiplied by $Q^2$, in logarithmic
scale up to $Q^2=10$ GeV$^2$. The different curves correspond to 
Eq.~(\protect\ref{a}) (solid), Eq.~(\protect\ref{b}) with
$\beta\rightarrow\infty$ (dotted), their upper limits  
 (long- and dot-dashed) and Eq.~(\protect\ref{c}) 
(dashed), respectively.
}\label{fig:ffpi}
\end{figure}

The present study mainly relies on analytic 
expressions of form factors \cite{AMGH2}. They have been obtained 
using simple phenomenological wave functions such as a Gaussian one employed 
in an earlier work \cite{ALLE} or a Hulth\'en one which may be more realistic:
\begin{equation}
F_1^G(Q^2)=\frac{(1-v^2)^2}{ \sqrt{1+v^2} }\;
\; \exp \left[- \frac{v^2}{1-v^2} \;\frac{m^2}{b^2}\right],
\hspace{1cm} {\rm with} \;\;v^2=\frac{Q^2}{4\,M^2_{\pi} +Q^2}\;,
\label{a}
\end{equation}
\begin{eqnarray}
\nonumber 
\lefteqn{F_1^H(Q^2)=\frac{(1-v^2)^2}{v}\; \; \frac{\alpha\; \beta\; 
(\alpha+ \beta)}{(\alpha- \beta)^2}} \\ && \qquad \times 
\left( \frac{\arctan[v\sqrt{m^2-\alpha^2}\,/\alpha]}{\sqrt{m^2-\alpha^2}}
-\frac{\arctan[2\,v\sqrt{m^2-\alpha^2}\,/(\alpha+\beta-v^2(\beta-\alpha))]
}{\sqrt{m^2-\alpha^2}}
\right.  \nonumber \\ &&  \qquad \,\left. + \,
 \frac{\arctan [v\,\sqrt{m^2-\beta^2}\,/\beta]}{\sqrt{m^2-\beta^2}}
- \frac{\arctan[2\,v\sqrt{m^2-\beta^2}\,/(\alpha+\beta+v^2(\beta-\alpha))]
}{\sqrt{m^2-\beta^2}}  \right).
\label{b}
\end{eqnarray}
Numerical results are presented in Fig. \ref{fig:ffpi}. 
They include form factors calculated with reasonable parameters 
for the two  wave functions. In both cases, upper limits 
that essentially correspond to a point pion are shown (denoted 
in the figure by the subscript $>$).  
As a guide, a curve which represents a good 
approximation to the form factor of Ref. \cite{ALLE}, which includes 
the re-scaling made by the authors, is also given:
\begin{equation}
F_1(Q^2)= \frac{1-\tilde{v}^2}{\sqrt{1+\tilde{v}^2}}\;,
\hspace{1cm} {\rm with} \;\;\tilde{v}^2=\frac{Q^2}{16\,m^2_q +Q^2} \;.
\label{c} 
\end{equation}

In all cases involving the correct definition of the velocity,  
Eqs. (\ref{a}, \ref{b}), we found that the squared charge radius 
has an incompressible value determined by the inverse of the 
squared pion mass, of the order of a few fm$^2$, one order 
of magnitude larger than experiment. This prevents one from making  
a sensible fit to the measured form factor (assuming we would like 
to do it). On the other hand, we cannot reproduce in any way its expected 
asymptotic power law, $Q^{-2}$. The fall off 
is always too fast. These results  are confirmed 
by using a more realistic but numerical wave function. 

\section{DISCUSSION AND CONCLUSION }
The present results  confirm those obtained in a theoretical model \cite{DESP}. 
They indicate that the implementation of the point-form approach used until 
now requires major improvements. These  ones include the contribution of 
two-body currents that could remove the main drawbacks of present 
calculations in relation with  the dependence of form factors  
on the ratio $Q/M_{\pi}$. Ultimately, the features in relation with the 
Goldstone boson nature of the pion, which concern in particular the 
asymptotic behavior and the charge radius, should be accounted for. 
The present  re-evaluation leaves room for them.

While developping this program, it may be useful to keep in mind 
other results. Recently, Coester and Riska \cite{COES} showed that 
the nucleon form factors could be successfully reproduced in the limit 
of a point nucleon. As this approximation  corresponds to our upper 
limits (denoted with the subscript $>$), it appears that what works 
in one case does not in the other. The difference is mainly due to 
the mass of the system under conside\-ra\-tion. On the other hand, results 
very similar to the present ones are obtained in other forms but for 
non-standard kinematics \cite{AMGH1,SIMU} that are known as giving 
incomplete results (see also field-theory motivated approaches 
\cite{BAKK,MELO}). The relationship may help to solve the problems 
evidenced by the implementation of the point-form approach used in this work.

\end{document}